\documentclass[reprint,aps,prl,superscriptaddress]{revtex4-1}
\pdfoutput=1
\usepackage{amsmath,bm}
\usepackage{epsfig}
\usepackage{setspace}
\usepackage{color}

\def\uu{{\bm u}}
\def\xx{{\bm x}}

\begin{document}
\newcommand{\aap}{Astron. Astrophys.}
\newcommand{\araa}{Annu. Rev. Aston. \& Astroph.}
\newcommand{\aapr}{Astron. Astrophys. Rev.}
\newcommand{\mnras}{Mon. Not. R. Astron. Soc.}
\newcommand{\apjl}{Astrophys. J. Lett.}
\newcommand{\apjs}{Astrophys. J. Suppl.}
\newcommand{\jfm}{J. Fluid Mech.}
\sloppy
\def\uu{{\bm u}}
\def\bb{{\bm b}}
\def\rr{{\bm r}}
\def\kk{{\bm k}}

\title{Dust-polarization Maps for Local Interstellar Turbulence}

\author{Alexei G. Kritsuk}\email{akritsuk@ucsd.edu}
\author{Raphael Flauger} \email{flauger@physics.ucsd.edu}
\affiliation{University of California, San Diego, 9500 Gilman Drive, La Jolla, California 92093-0424, USA }
\author{Sergey D. Ustyugov}
\affiliation{Keldysh Institute for Applied Mathematics, 4 Miusskaya Square, 125047, Moscow, Russia}

\begin{abstract} 
We show that simulations of magnetohydrodynamic turbulence in the multiphase interstellar medium yield an $E/B$ ratio for polarized emission from Galactic dust  in broad agreement with recent {\em Planck} measurements. In addition, the $B$-mode spectra display a scale dependence that is consistent  with observations over the range of scales resolved in the simulations. The simulations present an opportunity to understand the physical origin of the $E/B$ ratio and a starting point for more refined models of Galactic emission of use for both current and future cosmic microwave background experiments.
\end{abstract}


\maketitle

\paragraph{Introduction.}
Precision measurements of the polarization anisotropies of the cosmic microwave background (CMB) hold the potential to reveal deep insights into the physical process that generated the observed density perturbations (see, e.g., Ref.~\cite{kamionkowski.16} for a review). Several experiments are currently aiming to detect a polarization pattern on degree scales that is the characteristic signature of inflation. Others are about to join the search, and there is an active effort developing the plans for the next generation of experiments. 

From the current experiments, we already know that Galactic emission is brighter than the inflationary signal at all frequencies even in the cleanest patches of the sky (see Ref. \cite{dickinson16} for a review). As a consequence, a convincing detection requires exquisite control over Galactic foregrounds. Over the next decade, the sensitivity of experiments will improve by about 2 orders of magnitude so that the challenge posed by foregrounds will further increase.  

Existing data can provide useful insights, but the higher noise levels imply that the data cannot be used directly to prepare for the next generation of experiments. In this Letter, we show that {\em ab initio} simulations of magnetohydrodynamic (MHD) turbulence in the multiphase interstellar medium, together with a simple model for dust, lead to predictions that are in broad agreement with the scale dependence of the power spectra of $E$ and $B$ modes and the ratio of $E$- to $B$-mode power, and provide a promising way forward. 

\paragraph{Interstellar turbulence.---}%
The interstellar medium (ISM) is a complex magnetized mix of neutral and ionized gas, dust, cosmic rays, and radiation, all coupled through mass, energy, and momentum exchange. The conducting fluid component, including neutral and ionized hydrogen, is thermally unstable in certain density and temperature regimes \cite{field65} and tends to split into two or more stable thermal phases \cite{goldsmith..69,zeldovich.69}. The ISM is also constantly energized by supernova explosions and other relevant sources in the Galactic disk, which keep the fluid turbulent \cite{maclow.04,hennebelle.12}. Understanding the structure of this highly compressible multiphase magnetized turbulence is a challenge because of the multiscale nature of involved nonlinear interactions. Hence, the focus of recent studies was mostly on numerical experiments. These, in turn, concentrated primarily on MHD turbulence in isothermal fluids. The isothermal approximation, however, can only be valid within molecular clouds, where the gas temperature is about 10~K and would break on scales $\gtrsim1$~pc, where the presence of a warmer environment begins to play a crucial role. Therefore, isothermal simulations alone are of limited value for the discussion of large-scale ($\sim0.1-30$~pc) structure of interstellar turbulence probed by {\em Planck's} dust-polarization measurements.

The multiphase MHD turbulence simulations of Ref.~\cite{kritsuk..17} that we rely on here are complementary to the isothermal models, as they cover a range of scales $\sim0.5-200$~pc and include different coexisting MHD regimes of ISM turbulence. Under the local Galactic conditions, nonlinear relaxation leads to sub- or trans-Alfv\'enic conditions for the space-filling  warm and thermally unstable phases ($>$90\% by volume), while the cold phase  and molecular gas (comprising together $\sim$50\% by mass) remain super-Alfv\'enic \cite{heyer.12} \footnote{The turbulent Alfv\'enic Mach number  
${\cal M}_a\equiv\left(\overline{|\bm u|^2}/\overline{v_a^2}\right)^{1/2}$, where $v_a\equiv |\bm b|/\sqrt{4\pi\rho}$ is the Alfv\'en speed and the overbar denotes spatial average, is a proxy for the mean kinetic-to-magnetic energy ratio $K/M=4\pi\overline{\rho u^2}/\overline{b^2}\approx{\cal M}_a^2$}. The warm phase is transonic or slightly supersonic (relatively weak compressibility), while turbulence in the cold gas is hypersonic (very strong compressibility and abundance of shocks) \footnote{The turbulent sonic Mach number ${\cal M}_s\equiv\left(\overline{|\bm u|^2/c_s^2}\right)^{1/2}$, where $c_s$ is the speed of sound, measures the fluid compressibility.}. As a result, we observe an approximate kinetic-to-magnetic energy equipartition in the warm and unstable phases, and the fluid velocity aligns preferentially with the local magnetic field (Alfv\'en effect \cite{biskamp03}). This dynamic alignment is replaced by the kinematic alignment in the cold gas at high densities, where shocks are active and the kinetic energy dominates (i.e., the magnetic field aligns with one of the eigenvectors of a symmetric part of the rate-of-strain tensor \cite{tsinober09}). Moreover, at very high densities, where local gravitational instabilities  may lead to star formation, the velocity tends to align itself with the local gravitational acceleration (Zeldovich approximation \cite{zeldovich70,vergassola...94}), resulting in small-scale kinetic-to-gravitational potential energy equipartition. 

\begin{table*}[t]
\caption{\label{par}Model parameters.}
\footnotesize
\begin{tabular}{@{}ccccccccccccccc}
\hline
\hline
Case  & $b_0$ & $b_{\rm rms}$ & $b'_{\rm rms}$ & ${\cal M}_{a}$ & ${\cal M}_a^w$ & ${\cal M}_a^u$ & ${\cal M}_a^c$ & 
${\cal M}_{s}$ & ${\cal M}_s^w$ & ${\cal M}_s^u$ & ${\cal M}_s^c$ & ${\cal F}^w$ & ${\cal F}^u$ & ${\cal F}^c$ \\
     &    ($\mu$G) &($\mu$G) &($\mu$G) &&               &&&        &&&     &(\%)&(\%)& (\%)  \\
\hline
A  & 9.54 & 16.6 & 13.6 & 1.0 & 0.6 & 0.9 & 2.5 & 4.9 & 1.8 & 4.0 & 13.5  & 25 & 68 & 7    \\
B  & 3.02 & 11.7 & 11.3 &1.4 & 1.2 &1.6 & 4.3 & 5.4  & 1.7 & 4.2 & 15.2 & 23 & 70 & 7    \\
\hline
\end{tabular}
\\ \noindent
\parbox{17cm}{\raggedright
Input parameters: Box size $L=200$~pc, grid resolution $512^3$, mean field strength $b_0$~($\mu$G), mean density $n_0=5$~cm$^{-3}$, and rms velocity $u_{\rm rms}=16$~km/s. 
Statistically stationary conditions: rms field $b_{\rm rms}$, rms fluctuations $b'_{\rm rms}$, volume-averaged sonic (${\cal M}_s$) and Alfv\'enic  (${\cal M}_{a}$) Mach numbers, volume fractions of warm [$T>5250$~K, ${\cal F}_w$ (\%)], unstable [$184<T<5250$~K, ${\cal F}_u$ (\%)], and cold ($T<184$~K, ${\cal F}_c$ [\%]) thermal gas phases and corresponding phase-average Mach numbers ${\cal M}^{w,u,c}_{s,a}$.}
\end{table*}

The simulations also show that the probability density function (PDF) of the magnetic field strength is highly non-Gaussian \cite{kritsuk..17}. In both multiphase and isothermal simulations, the PDFs display fat extended stretched-exponential tails---a likely signature of strong spatial and temporal intermittency (locality of strong perturbations). The sites of strongest erratic field fluctuations are typically associated with filamentary dissipative structures formed by shocked cold and mostly molecular phase. In addition, in the relevant strongly magnetized cases, the PDF of magnetic fluctuations parallel to the mean field displays strong asymmetry and a flattened core \cite{kritsuk..17}.

As a consequence, simplified frameworks of weak MHD turbulence \cite{caldwell..17}, phenomenological models of the ISM \cite{ghosh+17}, or isothermal MHD turbulence \cite{king...17} may provide some insight but are not expected to convincingly explain observations and yield a robust understanding of current observations such as the $E/B$ ratio. At the same time, it is suggestive that MHD turbulence at Alfv\'enic Mach numbers ${\cal M}_a\lesssim0.5$ can provide $E/B$ ratios similar to those observed in the dust-polarization maps~\cite{kandel..17}. 

Simulations show that three key modes of self-organization, stimulating the dynamic, kinematic, or gravitational alignments, are consistent with the observed topology of polarization angles, tracing the plane-of-sky (POS) direction of the magnetic field with the column density and velocity centroid structures in dust- and synchrotron-polarization maps of local molecular clouds \cite{soler.....13,casanova.17,lazarian...17,soler.17}. 
The tangible success of multiphase models of the turbulent local ISM in interpreting the alignment properties of filamentary structures \cite{andre17} and recovering many other key observables \cite{kritsuk..17} suggests that the same models may perhaps also capture the observed $E/B$ ratios \cite{planckXXX16} in synthetic dust-polarization maps. We explore such a possibility in this Letter and show that the multiphase MHD turbulence simulations of Ref.~\cite{kritsuk..17} with a simple model for dust predict $E/B$ ratios comparable to those observed in {\em Planck}.

\paragraph{Numerical data.}
We use data from two simulations of interstellar turbulence of Ref.~\cite{kritsuk..17}, which mimic the local ISM conditions at the solar circle, to generate synthetic maps of thermal dust emission. These are MHD simulations of driven multiphase turbulence in a periodic domain of 200~pc on a side. The model includes a mean magnetic field, large-scale random solenoidal forcing \cite{kritsuk...10,padoan...16}, and volumetric cooling and heating; see Table~\ref{par} for a list of relevant parameters. The two cases A and B differ only by the strength of the mean magnetic field $\bm b_0$ and bracket a number of observables for the local ISM reasonably well, including (i) the overall hierarchical filamentary morphology of the molecular gas and the alignment of filaments with
respect to magnetic field lines, (ii) the volume and mass fractions of different thermal phases, (iii) the PDFs of column density and thermal pressure, (iv) the ratio of the turbulent magnetic field component versus the regular field, (v) the linewidth-size relationship for molecular clouds, as well as (vi) the low rates of star formation per free-fall time; see Ref.~\cite{kritsuk..17} for details. We use the data to generate synthetic dust-polarization maps, assuming a constant gas-to-dust ratio and perfect alignment of dust grains with magnetic field lines. For each case, we process a set of $\sim$70 data cubes evenly distributed in time over a period of $\sim$30~Myr of statistically stationary evolution. We use individual snapshots to generate sample polarization maps, while the power spectra reflect averages over all data cubes.

\begin{figure*}
	\centering
	\includegraphics[scale=.44]{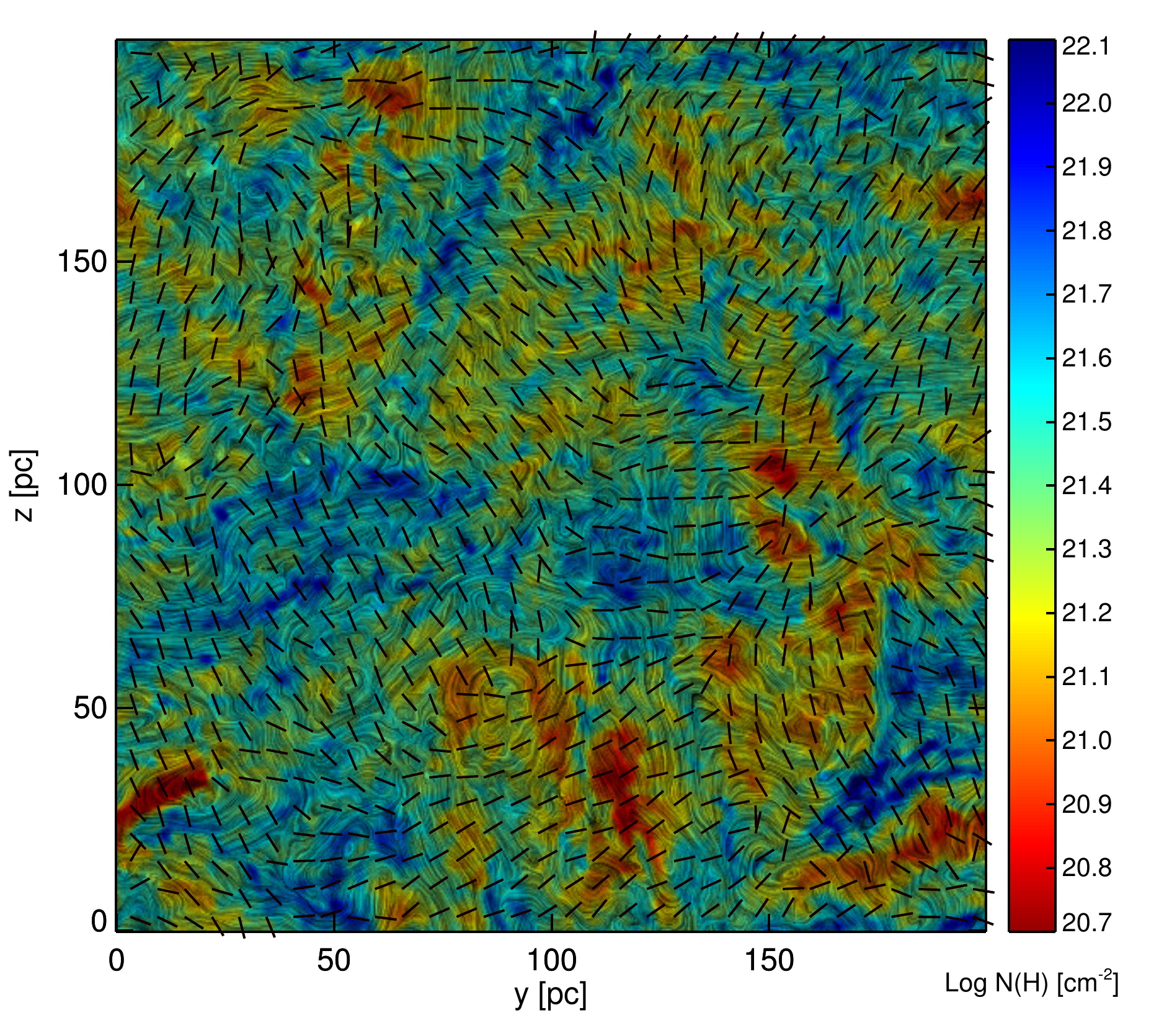}
	\includegraphics[scale=.44]{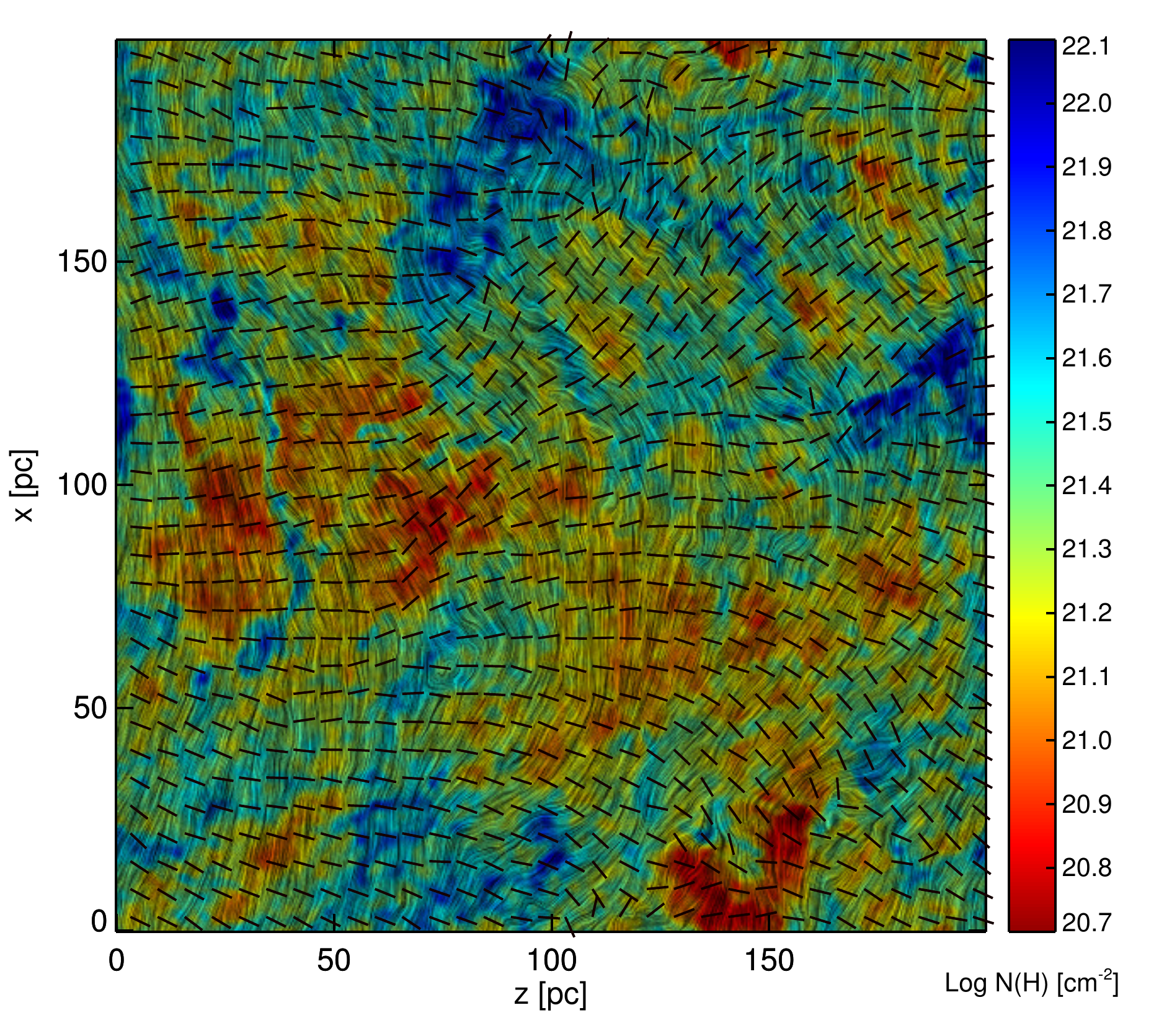}
	\caption{Sample synthetic polarization maps for case A corresponding to projections parallel (left) and perpendicular (right) to the mean field $\bm b_0$. The drapery texture generated using the line integral convolution (LIC) technique \cite{cabral.93} shows the POS magnetic field structure. Pseudovectors indicate the polarization direction (predominantly perpendicular to the field). Color shows the intensity in units of H{\sc i} column density. These maps are built on full-resolution numerical data smoothed with a low-pass boxcar filter of length 5 voxels, while no filtering is applied to compute the spectra. 
	}
	\label{map}
\end{figure*}

\paragraph{Polarization maps.}
For each grid point $\xx=(x,y,z)$, the data cubes contain a set of field values: $\rho(\xx)$, $\uu(\xx)$, $\bb(\xx)$, and $p(\xx)$---the fluid density, the velocity and magnetic field vectors, as well as the thermal pressure, respectively. The magnetic field vector $\bb=\bm b_0+\bb'$ includes the mean field $\bm b_0=(b_0,0,0)$ and fluctuations $\bb'$.

To construct the maps, we consider three line-of-sight directions, coinciding with principal coordinate axes and define projected quantities as functions of position {on the map $\bm r=(r_1,r_2)$, where $(r_1,r_2)$ can represent} $(x,y)$, $(y,z)$, or $(z,x)$ assuming polarized radiation is optically thin. For a projection along the $z$ direction, the intensity $I(\rr)$ and the Stokes parameters $Q(\rr)$ and $U(\rr)$ can be defined by
\begin{eqnarray}
&&\hskip 2.3cm I(\rr)\propto\int\rho(\rr,z) dz\,,\\
&&\hskip -.2cm Q\propto\!\!\int\!\!\epsilon(\rho){(b_y^2-b_x^2)}/{b^2}dz,\;\;\;
U\propto-2\!\!\int\!\!\epsilon(\rho) {b_xb_y}/{b^2}dz,
\end{eqnarray}
where $\epsilon(\rho)=\rho$ for dust grains that are perfectly aligned with the magnetic field. We will adopt a definition similar to the one used in Ref.~\cite{soler.....13} and set $\epsilon(\rho)=\rho\theta(\rho_t-\rho)$
controlled by the threshold density $\rho_t$ [here $\theta(\rho)$ is the Heaviside step function, and $\rho_t=\infty$ in the case of perfect alignment]. We will comment on the effects of this masking in more detail below. The polarized intensity is then given by $P=\sqrt{Q^2+U^2}$, while the polarization angle is $\psi_b=0.5\arctan\left({U}/{Q}\right)$. With these definitions, one can compute synthetic maps of scalar quantities $I(\rr)$, $Q(\rr)$, $U(\rr)$, and of a pseudovector $P(\rr)$. One can also compute the actual POS magnetic field for this same projection {$\tilde{\bm b}(\bm r)\equiv(\tilde{b}_1,\tilde{b}_2)=(\tilde{b}_x,\tilde{b}_y)$.}

As an illustration, Fig.~\ref{map} shows two sample maps for the strongly magnetized case A, using projections along axes parallel (left) and perpendicular (right) to the mean magnetic field. As expected, the thermal dust-polarization direction is mostly perpendicular to the direction of projected magnetic field $\tilde{\bm b}$ shown by the drapery pattern.  However, the two maps look qualitatively different overall, with a significantly more regular and anisotropic structure of $\tilde{\bm b}$ in the right panel, where the mean magnetic field lies in the POS.

\begin{figure*}
	\centering
	\includegraphics[scale=.65]{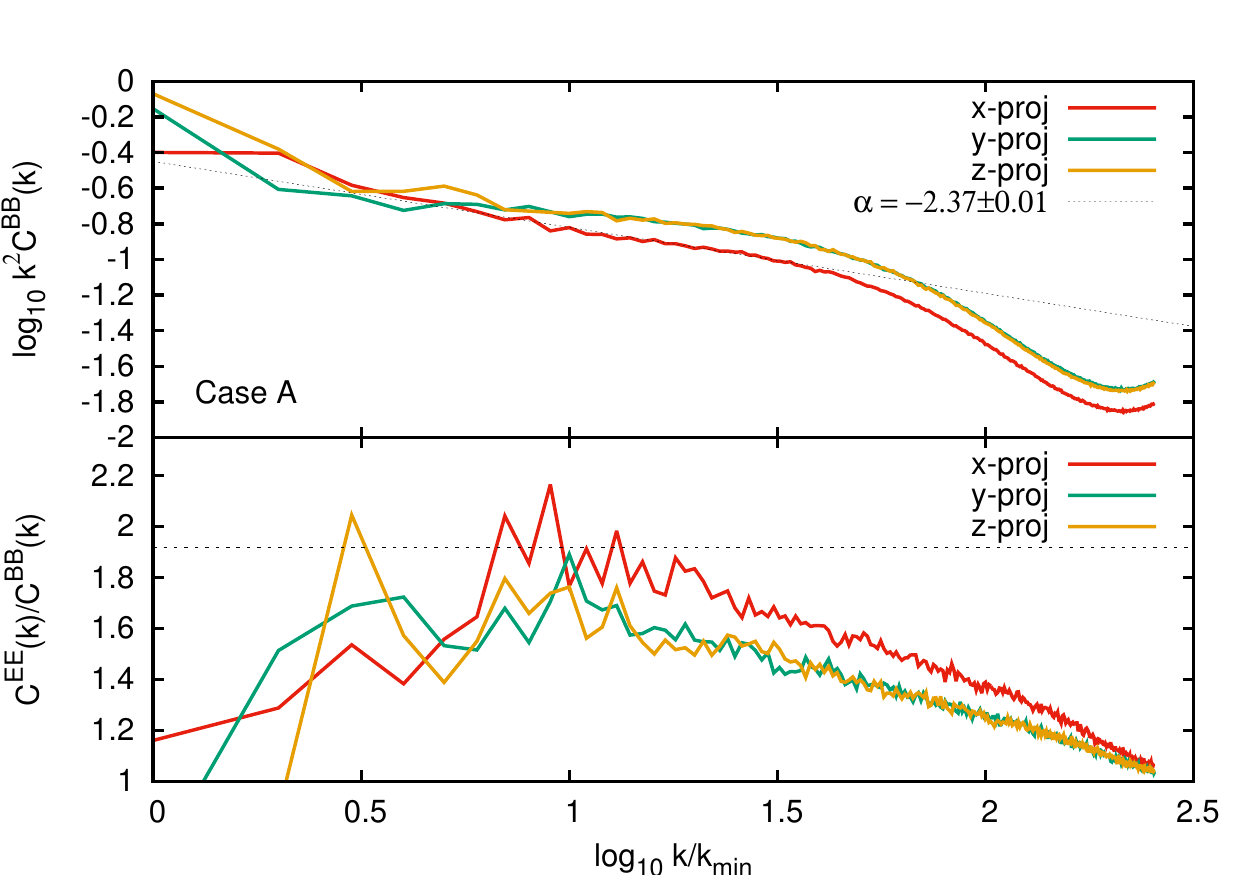}
	\includegraphics[scale=.65]{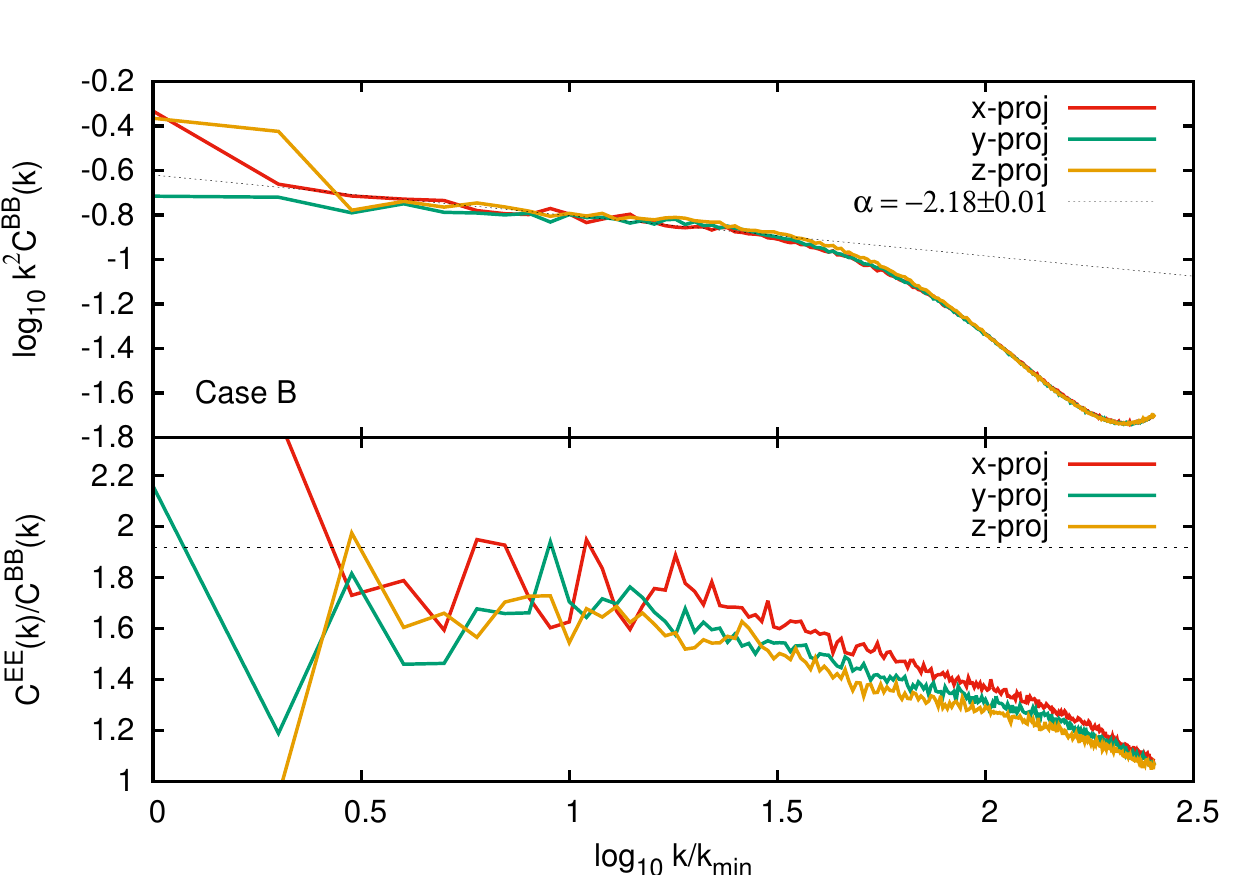}	
	\caption{ Time-average $B$-mode spectra in arbitrary units (top) and $E/B$ spectral ratios (bottom) for case A with strong large-scale magnetic field anisotropy (left) and less anisotropic case B (right). Dotted line in the top panels shows least-squares fit for projection along the mean field. Horizontal dotted line in the bottom panels indicates $E/B\approx1.92$ measured by {\em Planck} \cite{planckXXX16}. 
	}
	\label{pow}
\end{figure*}

\paragraph{Spectra.}
For each polarization map, we construct maps of the Fourier transforms of the $E$ and $B$ modes which are expressed using Fourier transforms of $Q(\rr)$ and $U(\rr)$, 
\begin{eqnarray}
\widehat{E}(\kk)&=&\frac{k_1^2- k_2^2}{k^2}\widehat{Q}(\kk)+\frac{2k_1k_2}{k^2}\widehat{U}(\kk),\\
\widehat{B}(\kk)&=&-\frac{2k_1k_2}{k^2}\widehat{Q}(\kk) +\frac{k_1^2-k_2^2}{k^2}\widehat{U}(\kk).
\end{eqnarray}
From these, we compute the $E$- and $B$-mode spectra defined by $C^{BB}(k)=\langle|\widehat{B}(\kk)|^2\rangle$ and $C^{EE}(k)=\langle|\widehat{E}(\kk)|^2\rangle$, taking an average $\langle\cdot\rangle$ over all wave vectors $\kk=(k_1,k_2)$ satisfying $|\kk|=k$. Figure~\ref{pow} shows the $B$-mode spectra averaged over the 70 realizations {with a power law scaling $C^{BB}(k)\propto k^{\alpha}$} (top), as well as the $E/B$ ratio (bottom) for all projections from cases A (left) and B (right). The calculations assume a threshold value $\rho_t=70$~cm$^{-3}$. 

The masked voxels comprise  (0.7--0.8)\% of the volume. The  spectra bear a signature of large-scale turbulence anisotropy in the strongly magnetized case A, with the $x$-projection spectrum carrying less power than those for the $y$ and $z$ projections. The spectra for different projections in case B are very similar since $|\bm b'|\gg b_0$ \cite{kritsuk..17}. 

Naturally, the spectra are subject to the usual resolution constraints implied by the numerics \cite{ustyugov...09,kritsuk+11}. The inertial range of scales, as usually defined, should be well separated from the forcing scale $k_f/k_{\rm min}\gg2$ and from the dissipation scale $k_{\eta}/k_{\rm min}\leq 30$. Hence, the spectral interval of interest here is limited to some range well within the interval of wave numbers $\log( k/k_{\rm min})\in[0.5,1.5]$.

In this range, $E/B\approx1.7$ for projections orthogonal to $\bm b_0$ in case A, while the parallel projection reaches somewhat higher levels $\approx$1.9. Case B demonstrates weaker anisotropy but similar levels of $E/B$. Overall, the $E/B$ ratios and spectral slopes  measured for $\rho_t=70$~cm$^{-3}$ are in broad agreement with {\em Planck} observations \cite{planckXXX16} in parts of the sky with comparable column densities. Whether there is statistically significant variation of the $E/B$ ratio in the {\em Planck} data reflecting the direction of local interstellar magnetic field \cite{zirnstein.....16} remains to be explored.

\paragraph{Masking.}
Since our simulations do not include self-gravity and the grid resolution is rather modest, the density range extends only to $\sim$10$^4$~cm$^{-3}$. 
Most of the masked gas with $\rho>70$~cm$^{-3}$ is not self-gravitating, as the Jeans-unstable volume fraction is $\sim$0.01\%, albeit most of the cold phase ($T<184$~K) is not masked, as ${\cal F}_c\approx7$\% \cite{kritsuk..17}. Thereby, applying the mask, we essentially remove a small part of mostly cold (presumably molecular \cite{padoan...16}), shocked supersonic, and super-Alfv\'enic material packed into dense filaments with the most erratic structure of the magnetic field that would otherwise contribute to the line-of-sight convolutions. {In case A, lower mask thresholds $\rho_t=50$ or 30~cm$^{-3}$ would result in $E/B\approx2.0$ or 2.2 and $\alpha\approx-2.46$ or $-2.59$, while affecting 1.4\% or 2.7\% of the domain volume, respectively.
At $\rho_t=70$~cm$^{-3}$, the mask effectively cuts off  stretched-exponential tails of the POS magnetic field PDF (not shown), leaving behind a compact exponential distribution.} The neglected 0.8\% would noticeably randomize synthetic dust-polarization maps, reducing the $E/B$ ratio to $\lesssim1.2$ and making the $B$-mode spectra shallower with {$\alpha\approx-1.6$}. This could be due to spurious numerical effects caused by the low grid resolution. Indeed, on a $512^3$ grid, the cooling timescale is not sufficiently resolved in the dense gas, causing artificial fragmentation of substructure within large-scale filaments. Higher-resolution models will show if masking is still needed. For simulations dedicated to CMB experiments that target regions of the sky with low column density, the mean density $n_0$ will be lower and will likely completely eliminate the need to  mask.

\paragraph{Caveats.}
In addition to numerics and model parametrization, there are several reasons to question the validity of physical assumptions we relied on to compute synthetic polarization maps. 
For instance, the efficiency of dust alignment with magnetic field by radiative torques (RAT) \cite{hoang.08} may change with the density, resulting in depolarization of thermal dust emission above some threshold. 
This may be caused by shocks that are known to actively shape the structure of dense regions in supersonic turbulence, resulting in a log-normal density PDF \cite{passot.98,kritsuk...07,kritsuk..17}. 
Shocks also tend to be preferred concentration sites for large (10~$\mu$m) dust grains \cite{tricco..17}. Since the RAT orientation mechanism is inefficient for small grains, collecting all larger grains at shocks would effectively reduce the alignment and result in depolarization.
Finally, recent studies indicate that the drag and Lorentz forces acting on the dust grains embedded in the turbulent ISM may strongly violate the usual assumption of constant dust-to-gas ratio \cite{hopkins.16,lee..17,monceau.17,tricco..17,hopkins.17}. Modeling small-scale dust segregation and size sorting in environments with realistic ISM turbulence and strong shocks would help to better inform dust-alignment models.

\paragraph{Conclusions and perspective.}
In this Letter, we have shown that MHD simulations of the turbulent, magnetized, multiphase ISM, assuming perfectly aligned dust grains and a constant dust-to-gas ratio, lead to predictions for the scale dependence of the angular power spectra of $E$- and $B$-mode polarization as well as the $E/B$ ratio that are broadly consistent with observations by {\em Planck}.

The 3D information available in MHD simulations allows us to incorporate realistic spatial variation of the properties of the dust and can be used to study the amount of decorrelation expected between different frequencies. Preliminary studies in this direction suggest that the amount of decorrelation is small. In addition, the simulations allow us to generate self-consistent maps of not only dust but also synchrotron emission. 

 This suggests that MHD simulations provide an opportunity to understand the physical conditions of the ISM from CMB data. In addition, the simulations furnish a promising starting point for the ISM modeling that can be used for the planning of future CMB experiments as well as tests of component separation techniques and analysis pipelines of existing experiments. 

\begin{acknowledgments}
A.K. appreciates discussions with Mike Norman and his sincere interest in the subject of this Letter and acknowledges significant contribution to this work by the late Dr. Sergey D. Ustyugov, who ran the simulations. This research was supported in part by the NASA ATP Grant No. 80NSSC18K0561.
  A.K. was supported in part by the National Science Foundation Grant No. AST-1412271. R.F. was supported in part by the Alfred P. Sloan Foundation and the Department of Energy under Grant No. DE-SC0009919. Computational resources were provided by the San Diego Supercomputer Center (Extreme Science and Engineering Discovery Environment Project No.~MCA07S014) and by the DOE Office of Science Innovative and Novel Computational Impact on Theory and Experiment (INCITE) program and Director's Discretionary grant allocated at the Oak Ridge Leadership Computing Facility, which is a DOE Office of Science User Facility supported under Contract DE-AC05-00OR22725  (Grants No. ast015 and No. ast021).
\end{acknowledgments}

\end{document}